\documentclass[runningheads]{llncs}

\usepackage{eccv}

\usepackage{booktabs}
\usepackage{graphicx}
\usepackage{caption}
\usepackage{subcaption}

\usepackage{booktabs}
\usepackage{graphicx}
\usepackage{caption}
\usepackage{subcaption}
\usepackage{float}

\usepackage{eccvabbrv}

\usepackage{graphicx}
\usepackage{booktabs}

\usepackage[accsupp]{axessibility}  
\usepackage{dirtree}
 \usepackage{array} 
 \usepackage{tabularx}
 \usepackage{graphicx}

\usepackage[pagebackref,breaklinks,colorlinks,citecolor=eccvblue]{hyperref}

\usepackage{orcidlink}

\begin{document}

\title{Civiverse: A Dataset for Analyzing User Engagement with Open-Source Text-to-Image Models} 

\titlerunning{ }

\author{Maria-Teresa De Rosa Palmini \and
Laura Wagner \and
Eva Cetinic}

\authorrunning{De Rosa Palmini et al.}

\institute{University of Zurich, Zurich, Switzerland}

\maketitle

\begin{abstract}

Text-to-image (TTI) systems, particularly those utilizing open-source frameworks, have become increasingly prevalent in the production of Artificial Intelligence (AI)-generated visuals. While existing literature has explored various problematic aspects of TTI technologies, such as bias in generated content, intellectual property concerns, and the reinforcement of harmful stereotypes, open-source TTI frameworks have not yet been systematically examined from a cultural perspective. This study addresses this gap by analyzing the CivitAI platform, a leading open-source platform dedicated to TTI AI. We introduce the \textbf{Civiverse} prompt dataset, encompassing millions of images and related metadata. We focus on prompt analysis, specifically examining the semantic characteristics of text prompts, as it is crucial for addressing societal issues related to generative technologies. This analysis provides insights into user intentions, preferences, and behaviors, which in turn shape the outputs of these models. Our findings reveal a predominant preference for generating explicit content, along with a focus on homogenization of semantic content. These insights underscore the need for further research into the perpetuation of misogyny, harmful stereotypes, and the uniformity of visual culture within these models.

\end{abstract}

\section{Introduction}
When OpenAI announced DALL·E, a TTI system with unprecedented capabilities \cite{radford2021learning}, and released the Contrastive Language and Image Pre-training (CLIP) vision transformer architecture \cite{radford2021learning}, it laid the foundation for its open-source competitors. Initially employing VQGAN \cite{esser_taming_2021} and later adopting diffusion models \cite{rombach_high-resolution_2022} with CLIP, open-source Text-to-Image (TTI) pipelines have consistently leveraged advancements in combined image and text embedding learning, achieving impressive results comparable to those of OpenAI. The release of Stable Diffusion in August 2022 \cite{noauthor_stable_nodate} significantly accelerated the development in open-source TTI, enabling the creation of images that are not only photorealistic but also aesthetically pleasing, from textual user input called prompts. TTI models are currently utilized in diverse sectors, with open-sourced Stable Diffusion models, currently accounting for the majority of AI-generated visuals, outpacing those that are API-restricted. \cite{noauthor_ai_2023}

Despite these advancements, extensive research has identified harmful content in multimodal TTI datasets \cite{birhane_multimodal_2021} concerning LAION-400M \cite{schuhmann2021laion} and even more problematic in its successor LAION-5B \cite{schuhmann2022laion} where Child Sexual Abuse Material (CSAM) was found, causing all LAION datasets to be temporarily taken down \cite{thiel_identifying_2023}. Additionally, cultural bias remains a significant issue. To be more specific, research has revealed that AI systems are predominantly trained using Western-centric data, resulting in the representation of the cultural values of Western, Educated, Industrialized, Rich, Democratic (WEIRD) societies \cite{masoud2023cultural}. This prevalence of WEIRD societies' values in AI systems perpetuates existing biases and stereotypes, particularly those associated with race, ethnicity, and gender \cite{bianchi2023easily, cho2023dall}. Moreover, issues arise from training on copyrighted material (e.g., LAION-5B dataset \cite{baio2022exploring}), the appropriation of particular artistic styles \cite{mccormack_no_2024}, the creation of misleading "deep-fake" images \cite{bird2023typology}, the risk of diminishing cultural and aesthetic diversity \cite{turk2023ai}, as well as the lack of authenticity in the generated content.

\begin{figure}[h]
    \centering
    \begin{minipage}[t]{0.45\textwidth}
        \centering
        \textbf{(a)}
        \begin{verbatim}
1girl, looking at viewer, face focus,
upper body, closed mouth, high definition,
1HEAD, best quality, ultra high res,
photorealistic, RAW photo,
physically-based rendering, shiny skin,
curvy, simple background.

        \end{verbatim}
    \end{minipage}
    \hfill
    \begin{minipage}[t]{0.45\textwidth}
        \centering
        \textbf{(b)} \\
        \includegraphics[width=0.5\textwidth]{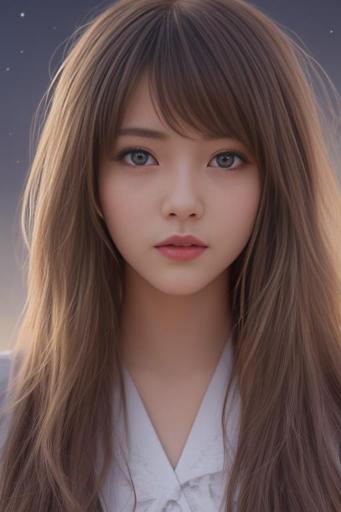}
    \end{minipage}
    \caption{A typical text-to-image prompt (a) and its associated image generation output (b).}
    \label{fig:text_image}
\end{figure}

The aforementioned issues do not always originate from the training data alone but can also be reflected and perpetuated through user-provided short text descriptions, known as "prompts" (example illustrated in Figure \ref{fig:text_image}). Based on a prompt, the model generates or "infers" output based on patterns extracted from the data it was trained on. TTI systems typically accept both a positive and a negative prompt as input. To be more specific, a positive prompt guides the diffusion model to generate images based on the given words, while a negative prompt directs the process away from specific meanings by applying negative weights (often - but not necessarily - used together with the positive prompt). \footnote{The usage of a positive prompt does not necessarily require a negative prompt to be used in conjunction. Also, not for every image the corresponding prompts are known as users have the option to upload images to CivitAI that were generated locally.}

The way users interact with these models can mirror the culture of emerging communities, offering a clearer understanding of the impact of generative AI on visual art and culture. Evaluating prompts can therefore help identify patterns and reveal how wording, both semantically and syntactically, influences outputs. Ultimately, this understanding serves as an intermediate step in designing generative AI systems that respect human diversity and complexity while ensuring technical proficiency and social responsibility.

To understand how users are engaging with open-source TTI models, we introduce \textbf{Civiverse 6M} \footnote{\url{https://huggingface.co/datasets/Civiverse/Civiverse-6M}}, a large-scale prompt dataset scraped from the open-source text-to-image platform CivitAI  \cite{noauthor_civitai_nodate} - a platform that not only offers text-to-image generation as as service, but has established itself as a platform facilitating the exchange of TTI models and adapters \footnote{Adapters specialize models for particular tasks or domains without necessity of retraining the entire model.}. The entire dataset consists of 6,546,165 image-URLs and the respective metadata spanning 7 months from October 2023 to April 2024. 

The \textbf{Civiverse 6M} dataset of generated imagery and respective metadata was the basis for the comprehensive analyis conducted in this paper. While taking other characteristics derived from the metadata into account, the primary focus of our study lies in the analysis of the text prompts, particularly examining their semantic characteristics. This study is aimed at identifying and addressing societal issues, specifically related to uncensored open source TTI pipelines and aims at providing a deeper understanding of user intentions, preferences, and behaviors.

\section {Background and Related Work}

Existing research has demonstrated that biases related to race, gender, and appearance propagate to multi-modal model outputs, affecting tasks like image captioning and image search. Image captioning models, for instance, tend to stereotype people based on gender or produce poorer quality captions for images of darker-skinned individuals \cite{zhao2021understanding}. Similarly, biases in image search algorithms reinforce negative stereotypes, impacting the sense of belonging for affected groups \cite{metaxa2021image}. The widespread use of image generation systems can thus have adverse effects on minority groups who are misrepresented or underrepresented. The aforementioned biases in TTI systems can originate from multiple sources, such as training data scraped from the web, which often contains harmful content and mislabeled examples. Filtering this data for "aesthetic" or "safe for work" content can introduce further biases, as observed with Dall·E 2's attempts to filter explicit content leading to bias amplification \cite{schuhmann2022laion}. However, they can also be perpetuated in the generated image through the text prompts that is used by users when interacting with TTI models. 

Given that, recent advancements in generative AI, particularly with diffusion models, explore the emerging culture of TTI models through the field of prompt engineering, specifically focusing on how human users interact with them through prompt manipulation. The DiffusionDB dataset \cite{wang2022diffusiondb}, for instance, provides a vast repository of 14 million images and 1.8 million unique prompts used with Stable Diffusion, shedding light on the syntactic and semantic aspects of prompts and their impact on model performance and errors. Concurrently, datasets such as ArtWhisperer \cite{vodrahalli2023artwhisperer} explore how users iteratively refine prompts to achieve desired outputs, drawing from a dataset of over 50,000 human-AI interactions to analyze prompt strategies and diversity. Moreover, research on prompt engineering in vision-language models \cite{gu2023systematic} has been comprehensive, covering methods, applications, and ethical considerations across text generation, image-text matching, and text-to-image models. Together, these studies deepen our understanding of prompt dynamics, user behavior, and the ongoing development of more effective and ethical AI systems.

In this study, inspired by previous work on prompt engineering by \cite{sanchez2023examining}, we adopt a novel approach to bias identification by extracting prompt specifiers based on frequency and mapping them to a high-dimensional 2D space. This method aims to provide a comprehensive understanding of the most popular topics and names, as well as user interactions with TTI models. By analyzing these interactions, we aim to better understand how emerging communities are using TTI models and work towards developing more effective and ethical AI systems.

\label{sec:intro}

\section{The \textbf{Civiverse 6M} Dataset}

\begin{figure} [h!]
    \centering
    \makebox[\textwidth][c]{\includegraphics[width=1\textwidth]{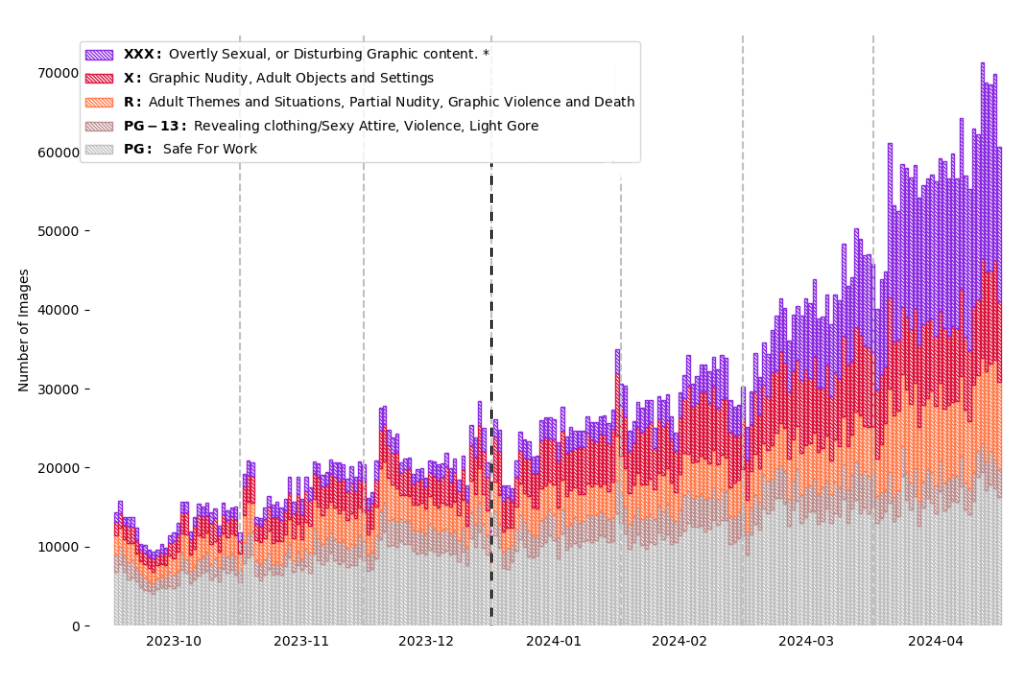}}
    \caption{Histogram daily uploaded images, spanning October 2023 - April 2024 color coded by their classification assigned by CivitAI. }
    \label{fig:civiverse}
\end{figure}

As a consequence of open source TTI advancements, various communities and platforms offering services related TTI synthesis have emerged. CivitAI, funded by A14z and launched in late 2022, offers on-site image generation as well as the exchanges of Stable Diffusion model derivatives. More than 6.5 million images have been posted from October 2023 to April 2024 (\Cref{fig:civiverse}). The platform is gaining popularity in a rapid pace: in September 2023, the average number of posted daily images was 13,429. On June 24 \textsuperscript{th} the number of uploaded daily images surpassed 120,000, which is an increase of almost tenfold in a span of eight months. The amount of generated imagery on the platform is expected to be much higher, as only the images generated and then deliberately posted by its users are available through the API. Using the CivitAI open REST API \cite{noauthor_rest_nodate}, we assembled a dataset, obtaining metadata for 6,546,165  images both created and published directly on the site, as well as those produced externally and uploaded. Notable is the high proportion of NSFW content observed in the data. The content rating system contrived by the CivitAI team \cite{civitai_civitais_2024} seems to be loosely based on the Motion Picture Association Film Rating system \cite{www.filmratings.com}, but tailored to the content of the site, ranging from PG (safe for all ages) to XXX (Overly Sexual or Disturbing Graphic content). The percentage of content rated above PG13 has risen from 55.19\% in October 2023 to 72.94 \% (\Cref{fig:civiverse}) in April 2024 in a span of 7 months.
\begin{table}[h!]
\centering
\begin{tabular}{|c|c|c|c|c|}
\hline
Year & Month & Total Images & Average Daily Images & NSFW Ratio \\ \hline
2023 & 10 & 405,920 & 13,014.10 & 0.5519 \\ \hline
2023 & 11 & 552,773 & 18,393.53 & 0.5829 \\ \hline
2023 & 12 & 661,746 & 21,310.68 & 0.5597 \\ \hline
2024 & 1 & 817,261 & 26,326.52 & 0.5859 \\ \hline
2024 & 2 & 926,644 & 31,902.31 & 0.6121 \\ \hline
2024 & 3 & 1,328,061 & 42,759.94 & 0.6504 \\ \hline
2024 & 4 & 1,853,760 & 61,476.00 & 0.7294 \\ \hline
\end{tabular}
\caption{Image data from October 2023 to April 2024}
\label{tab:image_data}
\end{table}

The \textbf{Civiverse 6M} holds the respective positive and negative prompt for each image, as well as other metadata. This includes identifiers such as the web URL pointing to the image, an image hash, a post id and a timestamp as well as the user name of the creator which we anonyomized using unique hashes. Additionally the metadata holds further parameters used for the conditioning beside the prompt (e.g. sample count, sampler type, cfg scale) being used in the generative process for the respective image. The name of the model being used, as well as the Variational Autoencoder (VAE) \cite{kingma_auto-encoding_2022} and model adapters such as LoRA  \cite{ryu_low-rank_2024} are also included in the metadata. Lastly, attributes used for the organization of the site, sorting and filtering (e.g. NSFW boolean, NSFW level) of the content as well as attributes related to community building features (e.g. like count) are included.

We make \textbf{Civiverse 6M} \footnote{\url{https://huggingface.co/datasets/Civiverse/Civiverse-6M}} available upon request with a cc-by-nc-4.0 license. Allowing users to use the dataset for academic purposes. In addition, we share our code \footnote{\url{https://github.com/a6088985/prompt\_analysis\_civiverse}} that collects, processes, and analyzes the images and prompts upon request.

\begin{table}[h!]
    \centering
	\caption{Overview of respective metadata attributes obtained for each image }
	\begin{tabularx}{\textwidth}{|l|X|}
		\hline
		\textbf{Category} & \textbf{Attributes} \\
		\hline
		Identifiers & Download URL, ID, PostID, Hash, Timestamp, Username \\
		\hline
		Classification & NSFW Boolean , NSFW Level, Browsing Level \\
		\hline
		Conditioning & Seed, Size, Steps, Sampler, CFG Scale, \textbf{Prompts} \\
		\hline
		\textbf{Prompts} & \textbf{Prompt (Positive), Prompt (Negative)} \\
		\hline
		Resources & Diffusion Model, VAE, Model Adapters: e.g. LoRA, LoHA, LoCon, Upscaling Model, Motion Module \\
		\hline
		Miscellaneous & Count of Social Reactions (Like, Cry etc.) comments \\
		\hline
	\end{tabularx}
\end{table}

\section{Methodology}
This section outlines the methods and tools used to perform a semantic analysis of user prompts on the CivitAI platform. The primary goal of this analysis is to explore the most common topics and styles present in user prompts, as well as the use of names, which can reveal patterns in user behavior and how TTI systems might perpetuate or amplify societal biases and other ethically dubious content.

\subsection{Topic Modeling}

Inspired by previous work on prompt analysis \cite{sanchez2023examining, mccormack_no_2024} using the DiffusionDB dataset \cite{wang2022diffusiondb}, our study focused on 6.1 million unique positive user prompts and 5.7 million unique negative user prompts from the \textbf{Civiverse 6M} dataset. Given the challenges posed by raw prompts—varying in format, length, and content—individual prompt specifiers were analyzed instead. Prompt specifiers are text fragments that specify desired characteristics of the image output. For instance, examples of specifiers in Figure \ref{fig:text_image} include "face focus," "high definition," "best quality," "shiny skin," "curvy," and "simple background."

To standardize our analysis, a methodology similar to Sanchez \cite{sanchez2023examining} was adopted, where prompts were divided using commas as separators, and specifiers appearing at least 500 times and not exceeding 35 characters were considered. This higher threshold, compared to the reference methodology's 100 occurrences, was set due to the larger dataset size of 6.1 million prompts compared to 1.8 million. Stopwords and specifiers from CivitAI’s revised labeling system, such as LoRa specifications and PonyDiffusion quality indications, were excluded due to their lack of semantic relevance. As a result, approximately 19,000 specifiers for positive prompts and 11,000 for negative prompts were obtained. The most frequently used specifiers and their occurrences in user prompts are presented in Table \ref{Table_1}.

\begin{table}[tb]
  \caption{Ranking of the popular prompt specifiers and their occurrences.}
  \label{Table_1}
  \centering
  \begin{subtable}[t]{0.45\textwidth}
    \centering
    \caption{15 most frequent specifiers for positive prompts.}
    \label{tab:positive_prompt}
    \begin{tabular}{@{}ll@{}}
      \toprule
      Specifier & {}\\
      \midrule
      masterpiece & 1.42\\
      best quality & 1.22\\
      solo & 0.74\\
      looking viewer & 0.65\\
      long hair & 0.53\\
      8k & 0.42\\
      realistic & 0.38\\
      highly detailed & 0.34\\
      smile & 0.30\\
      high quality & 0.28\\
      highres & 0.28\\
      blush & 0.26\\
      full body & 0.26\\
      large breasts & 0.25\\
      photorealistic & 0.25\\
      \bottomrule
    \end{tabular}
  \end{subtable}
  \hfill
  \begin{subtable}[t]{0.45\textwidth}
    \centering
    \caption{15 most frequent specifiers for negative prompts.}
    \label{tab:negative_prompt}
    \begin{tabular}{@{}ll@{}}
      \toprule
      Specifier & {Freq(\%)}\\
      \midrule
      bad anatomy & 2.01\\
      watermark & 1.85\\
      worst quality & 1.77\\
      blurry & 1.74\\
      low quality & 1.66\\
      text & 1.56\\
      signature & 1.33\\
      ugly & 1.30\\
      deformed & 1.26\\
      lowres & 1.19\\
      bad hands & 1.06\\
      monochrome & 0.97\\
      cropped & 0.85\\
      normal quality & 0.84\\
      extra limbs & 0.83\\
      \bottomrule
    \end{tabular}
  \end{subtable}
\end{table}

The MiniLMv2 model \cite{wang2020minilmv2} from the sentence-transformers library \cite{reimers-2019-sentence-bert} was used to transform prompt specifiers  into 384-dimensional embeddings. The Mini-LMv2, an optimized variant of MiniLM \cite{wang2020minilm}, was chosen for its ability to balance performance with reduced computational complexity, making it suitable for resource-limited environments. To analyze and visualize the prompt specifier embeddings, dimensionality reduction using UMAP (Uniform Manifold Approximation and Projection) \cite{mcinnes2018umap} was applied to map them into 2D points.

Subsequently, the HDBSCAN \cite{campello2013density} clustering algorithm was employed  to derive topics. HDBSCAN, known for its effectiveness in handling noisy data, especially when paired with UMAP \cite{asyaky2021improving, suryadjaja2021improving}, allowed for the extraction of cluster keywords by analyzing word distributions within each cluster. Additionally, class-specific term frequency-inverse document frequency (c-TF-IDF) was used to identify key terms within each group, accurately guiding the manual labeling of clusters.

\subsection{Most Popular Names}
\label{subsec:one}
The practice of utilizing artists' names in prompts to produce images in their style, commonly known as "style mimicry", has generated substantial concerns within the art community \cite{hoquet2023modern}. Models can replicate the artistic style of particular artists by fine-tuning on their work samples, infringing on copyrights and diminishing artists' motivation to create original pieces \cite{shan2023glaze}. This issue underscores the necessity of identifying the most frequently used artists' names in text-to-image prompts, as it can quantify the scale of style appropriation and inform strategies to safeguard artists' intellectual property and encourage their creative endeavors.

In this study, the SpaCy\footnote{\url{https://spacy.io/api/doc}} library was utilized to perform Named Entity Recognition (NER) with a focus on identifying PERSON entities, by employing the pre-trained NER model \textit{en\_core\_web\_sm}\footnote{\url{https://spacy.io/models/en}} provided by the library.

A temporal evolution analysis of the \textbf{Civiverse 6M} dataset was performed by splitting it into four distinct periods. The first period comprises September and October 2023 (with a total of 504,296 prompts), and the second period comprises November and December 2023 (with a total of 1,214,284 prompts). The analysis was conducted only on the positive prompts, as negative prompts typically do not contain named entities. The frequency of occurrences for the 10 most popular names per period was calculated to evaluate the evolution of these names over time. The frequency for each artist was computed as a percentage of the total prompts in their respective period.

\subsubsection{Compute Infrastructure.}
All experiments were conducted on a compute cluster running Debian GNU/Linux 10 with Intel Xeon E5-2650 v4 2.20GHz processors with 24 cores and 128GB RAM and Python 3.8.10.

\section{Results}

\subsection{Popular Topics}

\begin{figure}[h!]
    \centering
    \includegraphics[width=0.8\textwidth]{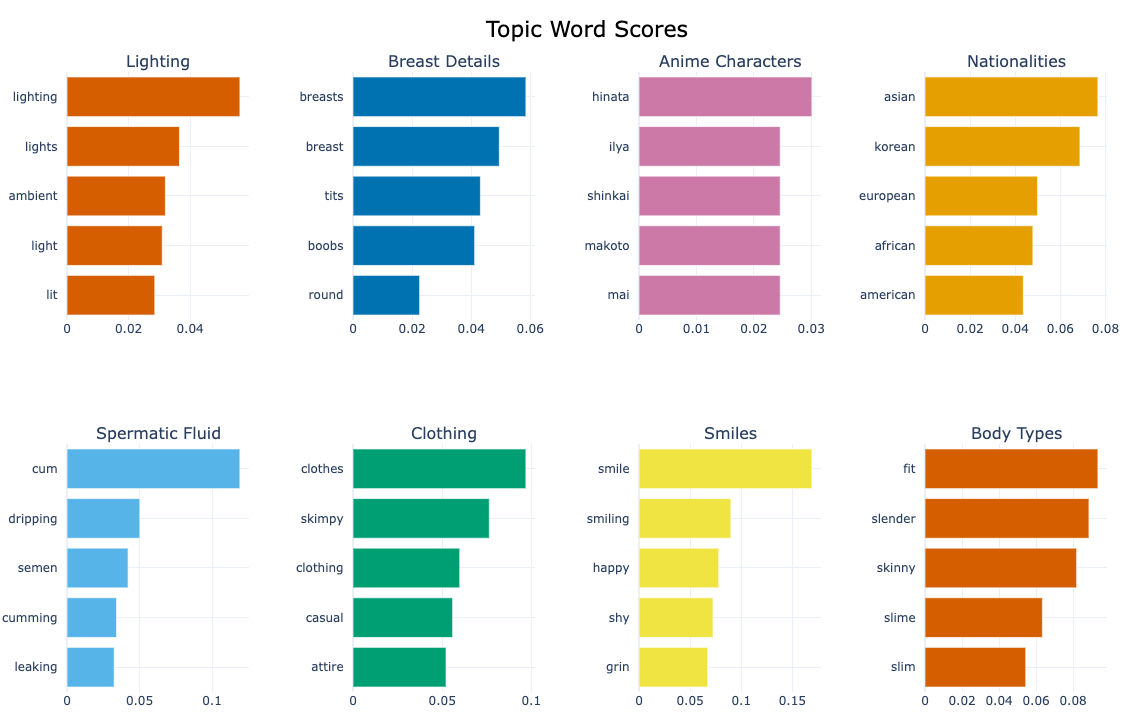}
    \caption{8 most popular topics for positive prompts of the Civiverse dataset.}
    \label{fig:topics_positive}
\end{figure}

\begin{figure}[H] 
    \centering
    \includegraphics[width=0.9\linewidth]{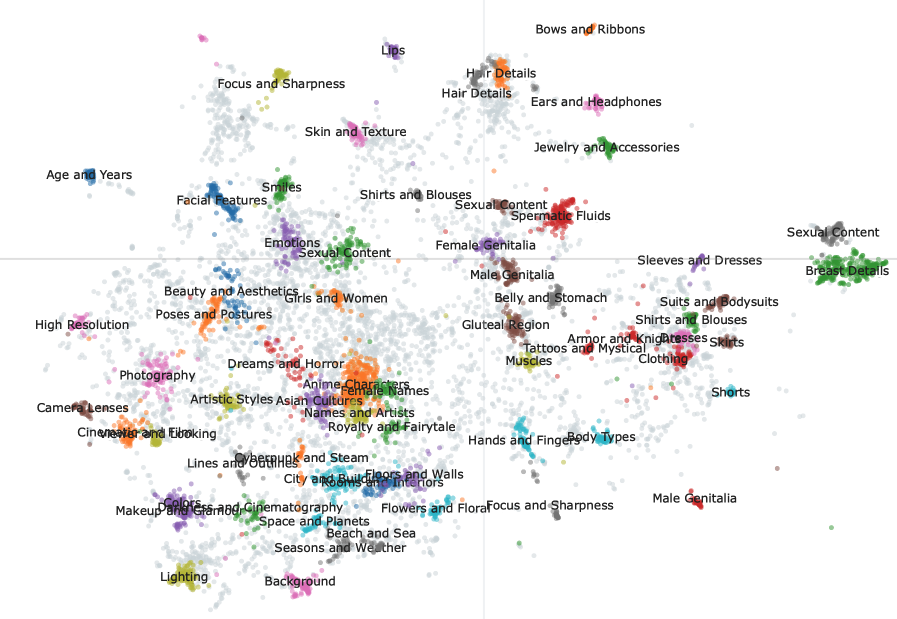}
    \caption{Visualization of positive prompt specifiers of the all-MiniLM-L6-v2 embeddings.}
    \label{fig:Visualization of positive prompt specifiers of the all-mpnet-base-v2 embeddings.}
\end{figure}

\textbf{(a) Positive Prompts:}
The results of the topic modeling on the 19,000 specifiers from the positive prompts yielded approximately 100 topics, with many being repeated or belonging to the same semantic category. In Figure \ref{fig:Visualization of positive prompt specifiers of the all-mpnet-base-v2 embeddings.}, we present the 50 most prominent thematic clusters. Additionally, a bar chart in Figure \ref{fig:topics_positive} displays the 8 most representative topics along with their corresponding keywords.

One of the primary categories in the categorization of specifiers is \textbf{Subject}. Within this category, the most frequently featured figures are humans, particularly women, with a notable prevalence of anime characters such as "Ochako Uraraka," "Inaho Kaizuka," and "Hinata." There is a discernible tendency towards specifying nationality, predominantly focusing on Asian cultures, especially "Japanese" and "Korean." Additionally, there is significant emphasis on \textit{Body Characteristics}, ranging from \textit{Body Types} (e.g., "fat girl," "slim waist," "fit figure") to \textit{Lower Body Parts} like \textit{Thighs} (e.g., "round thighs," "toned thighs," "juicy thighs"), \textit{Hips}, \textit{Arms} (e.g., "mechanical arm," "strong arms"), \textit{Feet} (e.g., "barefoot," "big feet"), and \textit{Belly} (e.g., "slim belly", "exposed belly"). Physical characteristics such as \textit{Skin Tone and Texture}, \textit{Hair Types}, and \textit{Smiles} (e.g., "friendly smile," "seductive grin") are also frequently specified.

There is also a noticeable tendency to emphasize \textbf{Sexuality and Nudity}, with detailed depictions of both \textit{Female and Male Genitalia}, female upper bodies, particularly \textit{Breasts}, and \textit{Male Spermatic Fluids}. Subjects are often portrayed in highly sexualized poses, described as "provocative," "alluring," and "sensual," with an aesthetic focus on body types using terms like "voluptuous hips" and "seductive physique."

Moreover, a notable tendency exists to specify the \textit{Point of View} towards the viewer, using phrases like "looking straight at the viewer," "smiling at the viewer," and "seducing the viewer," emphasizing direct engagement and making the viewer feel like a participant rather than an observer.

Specifiers also frequently indicate the \textbf{Medium} used in image generation, covering traditional and contemporary digital techniques. Photographic techniques are emphasized, with precise elements like \textit{Focus} (e.g., "manual focus"), \textit{ISO settings} (e.g., "iso100"), and various \textit{Camera Lenses} (e.g., "50mm"). High-resolution images are highlighted by terms like "hdr" and "4k." Additionally, Polaroid-inspired styles and filters (e.g., "polaroid," "vibrant") and \textit{Cinematic Aesthetics} (e.g., "cinematic," "hollywood") are prominent.

Furthermore, there is a tendency to include the \textbf{Names} of digital artists, photographers, and painters to prompt the model to imitate their style in the generated image. Section \ref{subsec:one} details the most frequently used names and their evolution over time.

Lastly, the categories of \textbf{Lighting} and \textbf{Colors} are crucial. Specifiers like "lighting," "shadows," "neon," and "glow" highlight the importance of ambiance and the dramatic impact of light and shadow. The Colors category shows a preference for rich, vibrant palettes, with specifiers like "vivid" and "vibrant" indicating a tendency towards bold, intense colors to enhance visual impact and convey a dynamic atmosphere.

\vspace{1cm}

\begin{figure}[h!]
    \centering
    \includegraphics[width=0.8\textwidth]{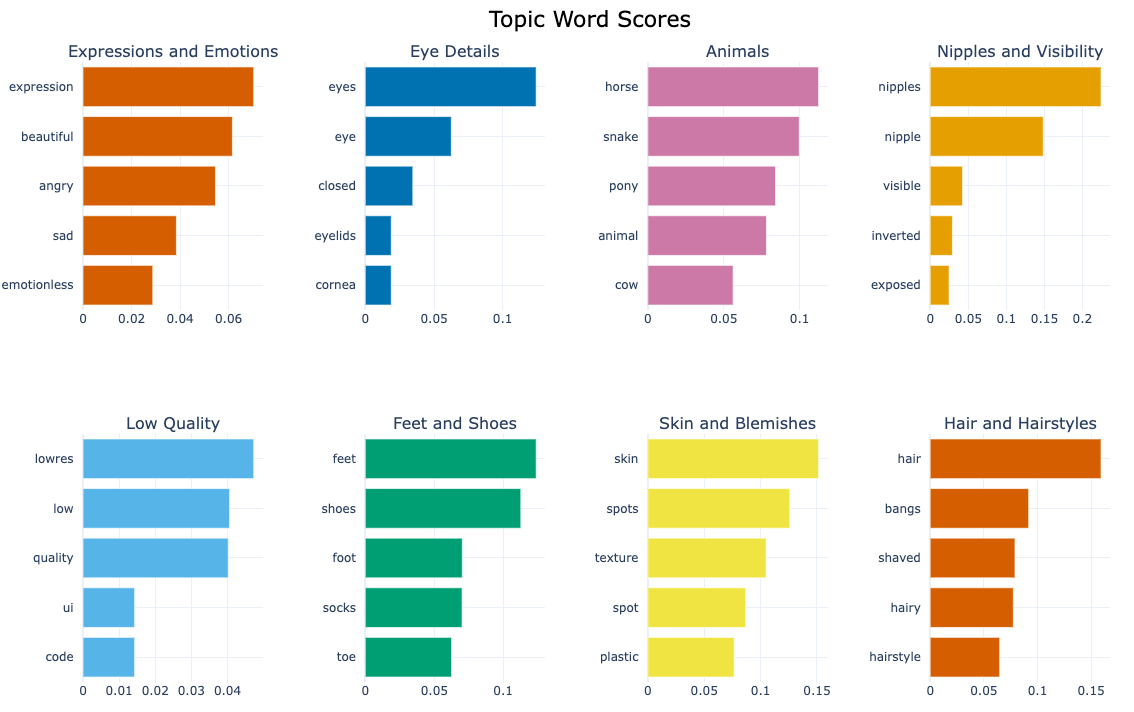}
    \caption{8 most popular topics for negative prompts of the Civiverse dataset.}
    \label{fig:topics}
\end{figure}

\begin{figure}[H] 
    \centering
    \includegraphics[width=0.9\linewidth]{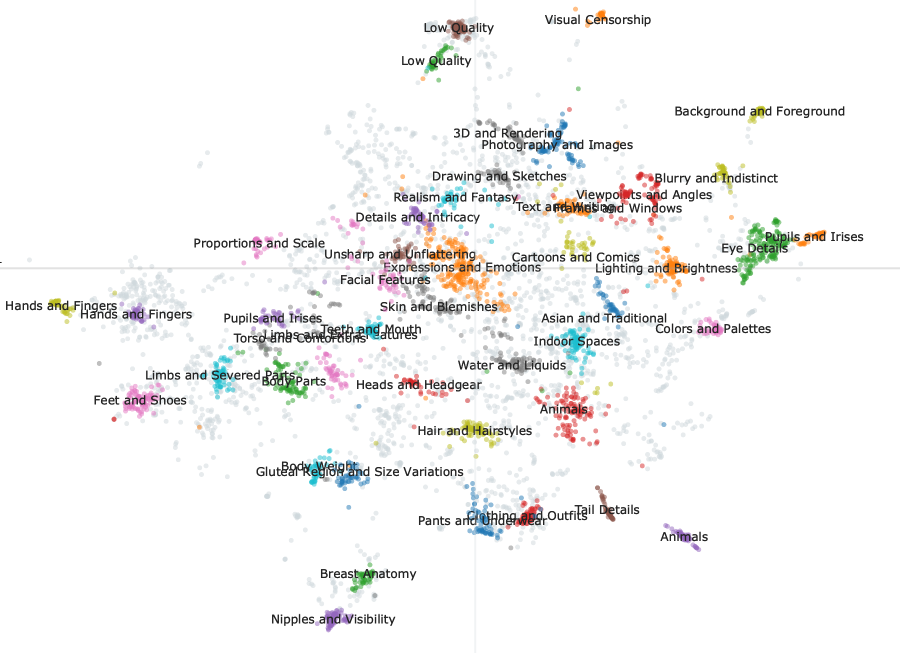}
    \caption{Visualization of negative prompt specifiers of the all-MiniLM-L6-v2 embeddings.}
    \label{fig:Visualization of negative prompt specifiers of the all-mpnet-base-v2 embeddings.}
\end{figure}

\textbf{(b) Negative Prompts:} 
Through the categorization of specifiers used for the negative prompts, we can gain a clearer understanding of which features and elements users prefer to avoid in their visually generated content. The results of the topic modeling on the 11,000 specifiers from the negative prompts yielded approximately 70 topics. In Figure \ref{fig:Visualization of negative prompt specifiers of the all-mpnet-base-v2 embeddings.}, we present the 50 most prominent thematic clusters. Additionally, a bar chart in Figure \ref{fig:topics} displays the 8 most representative topics along with their corresponding keywords.

As with positive prompts, the \textbf{Subject} category remains primary in negative prompts, particularly focusing on \textit{Expressions and Emotions}. Users generally avoid unsettling emotions such as "emotionless," "repulsive feeling," "disgusting look," or "bizarre," and unpleasant expressions like "overly angry," "extremely sad," or "mean face." Additionally, \textit{Facial Features} are frequently specified in negative prompts, including both broad descriptions (e.g., "weird," "bad," or "incomplete facial features") and more detailed ones. Specifically, undesirable \textit{Eye Features} (e.g., "closed eyelids," "under-eye bags," "misshaped eyes"), unattractive \textit{Pupils and Irises} (e.g., "deformed pupils," "distorted irises"), and unappealing \textit{Teeth and Mouth} details (e.g., "discolored," "missing teeth") are noted. Poor \textit{Skin Conditions}, such as "acne," "pimples," "noisy skin texture," and "age spots," are also commonly specified.

In addition to facial and emotional aspects, users frequently mention specific \textit{Animals} and their corresponding \textit{Animal Features}, likely aiming to avoid these elements in generated content. For example, animals like "spiders," "reptiles," and "wolves" are often mentioned negatively. Features such as "snake noses," "rabbit ears," "snake tongue," "cow faces," and "horse heads" are included to prevent unnatural or off-putting anthropomorphic characteristics.

Regarding \textit{Clothing}, users specify styles and types of attire to avoid in their negative prompts, such as "trashy," "extra," "maternity," "prisoner," "poor," and "badly fitted clothes." Certain types of clothing that are not stereotypically considered non-appealing, like "evening gowns," "fluttering costumes," and "shiny clothes," are also avoided to ensure the generated content aligns with the user's vision.

A significant number of negative prompts focus on the category of \textbf{Body Horror}, often depicting clear representations of the human body to counteract typical AI-generated imperfections. This category includes aspects related to \textbf{Sexuality and Nudity}, aiming to counteract unwanted representations of \textit{Female and Male Genitalia}, \textit{Female Breasts and Nipples}, and \textit{Gluteal Regions}. Specifications often address details related to shape and size, using terms like "small," "tiny," "medium," "huge," and "giant," as well as malformations and abnormalities such as "missing," "double," "misaligned," "multiple," and "weird." Additionally, colors to avoid, such as "colorful," "black," "red," and "green," are specified. Other aspects include avoiding disjointed or extra body parts, \textit{Severed Limbs}, \textit{Contorted Torsos}, and \textit{Malformed Necks and Collars}, as well as undesirable \textit{Body Weight} attributes like "overweight," "anorexic," and "fat woman."

In terms of \textbf{Resolution and Clarity}, users aim to counteract issues like \textit{Low Resolution}, which results in a lack of detail and sharpness, through specifiers such as "lowres" and "low quality." Issues related to \textit{Compression and Artifacts} (e.g., "jpeg large artifacts," "compressed images") are also addressed. For the \textbf{Lighting and Illumination} category, users avoid undesirable conditions with specifiers like "harsh light," "inconsistent light," and "neon light," as well as issues related to \textit{Shadows and Shading} (e.g., "heavy shadows," "flat shading"). In the \textbf{Color Palette} category, specifiers address unwanted color conditions, specifically targeting a lack of vibrancy (e.g., "missing colors," "messy colors," "faded colors," "dull colors," "unsaturated colors"), and issues with \textit{Saturation and Exposure} (e.g., "highly saturated," "oversaturation," "saturated colors").

\subsection{Popular Names}
The temporal analysis of the \textbf{Civiverse} dataset highlights significant trends in the use of artist names in TTI prompts, illustrating the practice of "style mimicry". Greg Rutkowski's name consistently appeared most frequently, indicating a strong and growing preference for his digital illustrations. Similarly, artists like Carne Griffiths, Luis Royo, and Frank Frazetta were prominently featured, each representing distinct artistic mediums such as media art, illustration, and cartooning. The inclusion of classical artists like Rembrandt underscores the ongoing appropriation of both contemporary and historical styles, reflecting a broad spectrum of influences.

\begin{table}[h!]
\centering
\caption{Top artists' names in prompts for Sep '23 to Dec '23}
\label{tab:table_3}
\begin{tabular}{llr|llr}
\toprule
\textbf{Sep \& Oct '23} & \textbf{Freq} & \textbf{Medium} & \textbf{Nov \& Dec '23} & \textbf{Freq} & \textbf{Medium} \\
\midrule
\textbf{Greg Rutkowski} & 0.78\% & Dig. Illustrator & \textbf{Greg Rutkowski} & 1.12\% & Dig. Illustrator \\
Charles Dwyer & 0.17\% & Painter & \textbf{Luis Royo} & 0.45\% & Illustrator \\
Daniel F. Gerhartz & 0.16\% & Painter  & \textbf{Carne Griffiths} & 0.38\% & Media Artist \\
J.C. Leyendecker & 0.16\% & Illustrator & \textbf{Rembrandt} & 0.36\% & Painter \\
Kuvshinov & 0.14\% & Digital Artist & Ross Tran & 0.28\% & Digital Artist \\
Flora Borsi & 0.13\% & Photographer & Enki Bilal & 0.21\% & Graphic Novelist \\
Ruan Jia & 0.12\% & Digital Artist & Michael Garmash & 0.21\% & Painter \\
\textbf{Frank Frazetta} & 0.10\% & Illustrator & \textbf{Steve McCurry} & 0.21\% & Photographer \\
Makoto Shinkai & 0.10\% & Filmmaker & Kiguri & 0.21\% & Dig. Artist \\
Pino Daeni & 0.10\% & Painter & Maria Sibylla Merian & 0.16\% & Sc. Illustrator \\
\bottomrule
\end{tabular}
\end{table}

\begin{table}[h!]
\centering
\caption{Top artists' names in prompts for Jan '24 to Apr '24}
\label{tab:table_4}
\begin{tabular}{llr|llr}
\toprule
\textbf{Jan \& Feb '24} & \textbf{Freq} & \textbf{Medium} & \textbf{Mar \& Apr '24} & \textbf{Freq} & \textbf{Medium} \\
\midrule
\textbf{Greg Rutkowski} & 0.68\% & Dig. Illustrator & Personalami & 0.52\%  & Dig. Artist \\
\textbf{Rembrandt} & 0.41\% & Painter & \textbf{Greg Rutkowski} & 0.48\% & Dig. Illustrator \\
\textbf{Carne Griffiths}  & 0.38\% & Media Artist & \textbf{Carne Griffiths} & 0.23\% & Media Artist \\
\textbf{Frank Frazetta} & 0.31\% & Cartoonist & \textbf{Rembrandt} & 0.20\% & Painter \\
\textbf{Luis Royo}  & 0.32\% & Illustrator & Nora Higuma & 0.11\% & Manga Artist \\
Johnfoxart  & 0.25\% & Dig. Artist & Alice Pasquini & 0.11\% & Street Artist \\
Thomas Kinkade & 0.20\% & Painter & Don McCullin & 0.11\% & Photojournalist \\
Don McCullin & 0.19\% & Photojournalist & Alphonse Mucha & 0.10\% & Painter \\
Moebius & 0.14\% & Cartoonist & \textbf{Steve McCurry} & 0.10\% & Photographer \\
Frederic Edwin Church & 0.14\% & Painter & Minjae Lee & 0.10\% & Portrait Artist \\
\bottomrule
\end{tabular}
\end{table}

Despite the prominence of these names, they still represent a small fraction of the overall number of prompts, suggesting that while certain artists' styles are popular, they are not overwhelmingly dominant in the dataset. This analysis also reveals a notable gender imbalance, with male artists being predominantly featured. Tables \ref{tab:table_3} and \ref{tab:table_4} present the 10 most popular names, the frequency of occurrence as a percentage, as well as the profession of the names involved. These findings provide therefore crucial quantitative insights into the scale of style appropriation, which is essential for developing strategies to protect artists' intellectual property and encourage original creative endeavors in the digital age.

\section{Conclusion}

In exploring the problematic aspects of TTI models, the emphasis has so far been on examining the visual outputs of popular TTI models or training data used for the development of such models. Our study shifts the focus by accumulating user input (prompts) alongside other metadata to better understand user interactions with TTI models. This approach reveals how these interactions can reinforce or challenge existing issues, providing insights into the broader societal implications of these technologies in an open-source setting. The trends identified in our analysis show that as the popularity and user base of platforms such as CivitAI and the hosted models grow, there is not only an absolute increase in pornographic content but also a disproportionate rise in its proportion relative to SFW content. The predominant and growing preference among the users for generating explicit content, and the homogenization of semantic content revealed a necessity for further research into the perpetuation of misogyny, harmful stereotypes, and the uniformity of visual culture that emerges from the interaction with models on platforms such as CivitAI. This initial analysis of the dataset was limited on linguistically evaluating user prompts (positive, negative), yet other attributes such as utilized models and adapters as well as parameters involved in the image genesis, will be included in a follow-up research to obtain a more granular understanding. 

Even though CivitAI is the most established model-sharing and open-source platform for image generation, the results might be skewed by the specific culture of its presumably presumably technologically adept user base and the fact that other established TTI services such as DALL·E or Adobe Firefly have strict policies on NSFW content. This may cause users with certain inclinations to gravitate towards open-source TTI alternatives. Therefore, the data might not reflect how open-source TTI AI is used by the entirety of TTI users across all systems. Nonetheless, platform usage statistics gathered over the past months show an accelerating growth rate, indicating that the platform is no longer limited to a small, specialized group.

\clearpage  

%
%
\bibliographystyle{splncs04}
\bibliography{main}
\end{document}